\documentclass[12pt,preprint]{aastex}
\usepackage{latexsym,amsmath,graphicx}

\begin{document}
\title{Surface temperature and synthetic spectral energy distributions for rotationally deformed stars}
\author{C.C. Lovekin, R.G. Deupree \& C.I. Short}
\affil{Institute for Computational Astrophysics and Department of Astronomy and Physics, St. Mary's University, Halifax, NS 
}

\begin{abstract}

Extreme deformation of a stellar surface, such as that produced by rapid 
rotation, causes the surface temperature and gravity 
to vary significantly with latitude.  
Thus, the spectral energy distribution (SED) of a non-spherical star could
differ significantly from the SED of a spherical star with the same average 
temperature and luminosity.  
Calculation of the SED of a deformed star is often
approximated as a composite of several spectra, each produced by a plane 
parallel model of given effective temperature and gravity.
The 
weighting of these spectra over the stellar surface, and hence the inferred 
effective temperature
and luminosity, will be dependent on the inclination of the rotation axis of
the star with respect to the observer, as well as the temperature and gravity 
distribution on the stellar surface.  Here we calculate the surface conditions
of rapidly rotating stars with a 2D stellar structure and evolution code and 
compare the effective temperature distribution
to that predicted by von Zeipel's law.

We calculate the 
composite spectrum for a deformed star by interpolating within a grid of 
intensity spectra of plane parallel
model atmospheres and integrating over the surface of the star.  This allows 
us to examine
the SED for effects of inclination and degree of deformation based on the 2D 
models.  Using this method, we find that the deduced variation 
of effective temperature with 
inclination can be as much as 3000 K for an early B star, depending on the 
details of the 
underlying model.  

As a test case for our models, we examine the rapidly rotating star Achernar
($\alpha$ Eri,
HD10144).  Recent interferometric observations have determined the star to be
quite oblate.  Combined with the ultraviolet SED measured by the OAO-2 
satellite, we are able to make direct comparisons with 
observations.

\end{abstract}



\section{Introduction}

Despite much effort, the structure and evolution
of rapidly rotating stars remains one of the major problems in
stellar theory for a number of theoretical and
observational reasons. One significant issue is the translation of
observational measurements (e.g., flux) into quantities that are related to 
the structure of the star, such as effective temperature and surface gravity.
This translation is a comparatively simple task for spherically symmetric 
stars, for which
the effective temperature and gravity are unique and their relationship with 
luminosity is well defined, but becomes more
difficult for rotating stars, where both the effective temperature and 
gravity vary over the
surface of the star. 
According to von Zeipel's law \citep{vonz}, the local radiative flux is 
proportional to
the local effective gravity, which is the sum of the force of gravity and the 
centrifugal force, so T$_{eff}$ $\propto$ g$^{1/4}_{eff}$.  This means
the spectrum of a rapidly rotating star is often treated as a 
composite of several spectra, each at a specific effective temperature and
gravity. 
Because the observed composite spectrum will vary with inclination, the 
values derived from observations will also depend on the inclination of the 
rotation axis with respect to the observer, something that is not known {\it a 
priori}.

A corollary of this law is that the relationship between
the luminosity and an observed bolometric magnitude also depends on the
inclination, as the amount of energy radiated from the 
stellar surface also varies with latitude. Thus, even something as relatively 
straightforward as assigning a
star's location in the HR diagram is not simple for rotating stars, as
the location would not be a point but a curve with inclination as the free
parameter. The length and shape of the
curve would depend on the amount of surface rotation and perhaps the
angular momentum distribution.  The inclination determines where on the
curve the observer would place the star.  This effect has been well studied
\citep{col66,hard68,maed70}.  These previous studies were done using 
spherical, uniformly rotating structural models.  The distortion of the 
surface was described using a
Roche potential and the surface variation in temperature followed von
Zeipel's (1924) gravity darkening law, with the total luminosity obeying
an equation of
the form L($\omega$) = L(0)f($\omega$), where $\omega$ is the fraction of
critical rotation ($\Omega$/$\Omega_{crit}$).  Differential rotation has been 
studied by \citet{col85}, using a cylindrical rotation law applied to A stars.
The interiors of these stars were modeled as for a 1D stellar model with 
three correction factors applied to account for the differential rotation.
They applied these models to produce synthetic photometric observations of
groups of stars.  In agreement with previous studies, they find that rotation
shifts a star's location in the HR diagram, and differential rotation results
in a larger shift.
In this paper, we take these models one step further, applying a fully implicit
2D stellar evolution code, described in \S \ref{rotorc},
with arbitrary rotation laws to produce our interior models.
These 2D evolution simulations allow us to assess how realistic von Zeipel's 
law is in several situations.  This will be examined in more detail in 
section \S \ref{sec:roche}.

The observed spectral energy distribution (SED) can be found from the weighted
sum of the radiative intensities emitted in the direction of the observer, 
integrated over the surface of the star.
In principle, the SED contains information about the angular variation of
the quantities which influence the radiation field.  In this paper, we 
examine what information, if any, can be determined about the angular 
momentum distribution, and hence the structure, based on the SED of a star.
We chose here to work with the SED rather than individual lines, as was done 
by \citet{col74} and \citet{col79} because we hoped to be able to employ this 
method as a general technique over a wide range of stars.  We also hoped to 
avoid dependence on the properties of any particular set of lines.

There are three numerical modeling components in this process. The
first is the calculation of fully 2-D stellar evolution sequences with
rotation \citep{bob90,bob95,bob98} to obtain the effective temperatures and
effective surface gravities as functions of latitude for any point in a
stellar evolution sequence. Here the effective
temperature is defined as the black body temperature required to produce the 
local surface flux, and the effective gravity is the component of the
centrifugal force and the gradient of the gravitational potential in the 
direction of the normal to the local surface (i.e., the local vertical). 
These two quantities are required as input parameters to the stellar 
atmosphere calculations.  We
assume that we can model the atmosphere at any given location on the
surface as a plane parallel atmosphere with this local effective
temperature and effective gravity. 
This approximation is good if the 
horizontal photon mean free path is very small compared to the horizontal distance
over which there are significant structural changes along the stellar
surface. The region where this approximation is least reliable is near the 
equator,
where the effective gravity is smallest and the horizontal structural
gradients the largest. This error is somewhat balanced by the fact that the 
equatorial 
region has the lowest effective temperature, so it contributes less to the 
observed flux except at inclinations of nearly ninety degrees.

The second modeling component is the calculation of a grid of
model atmospheres that cover the range of effective temperatures and
effective gravities required. For this we use the {\tt PHOENIX} model
atmosphere code \citep{hauschildt_b99}. 
These models are used to calculate the emergent intensities as a function of 
angle from the vertical, which will be integrated to obtain the observed flux.
The main advantage to this code is the ability to model many of the important
lines in non-local thermodynamic equilibrium (NLTE), while most previous 
studies (e.~g.~\citet{maed70}) used only LTE calculations.  

The third modeling component is the numerical integration of these
intensities to obtain the observed flux. The procedure used is very
similar to that described in \citet{cass87,lin94,town04}.
The surface of the star
is divided into a mesh in longitude and latitude. For a given
inclination the direction to the observer can be calculated at any
location on the surface and the appropriate intensity selected from the
input supplied by the model atmosphere code. This will be multiplied by
the local surface area element and the cosine of the angle between the local
surface normal and the direction to the observer. 
The sum of the intensities produced by all the mesh zones gives the spectral 
energy distribution of the star at a given inclination.
Before integration, the intensities are convolved to match the profile of the
OAO-2 spectrometers.  This profile covers too large a wavelength range (10-20 
$\mbox{\AA}$) for Doppler shifts to be noticeable.
For this reason, the Doppler shift has been neglected in this integration.

We chose this study 
because of the recent work of \citet{dom03}, showing that the sometime Be star
$\alpha$ Eri (Achernar, HD10144) is highly oblate based on optical 
interferometric observations, with an axial ratio a/b = 1.56 $\pm$ 0.05.
This oblateness, defined by the axial ratio, 
is determined by fitting an ellipse to a uniform disk model.  
The star is known to be relatively rapidly rotating, with vsin{\it i} = 
225 km s$^{-1}$ \citep{slet82}, but there was no indication from this or any 
other observations 
that the star was as oblate as indicated by the observations of \citet{dom03}.
We were interested to see if there were indications of the degree of deformation
in other available data.  If so, we might be able to use these other 
indications to be able to determine how prevalent highly oblate stars might be,
particularly for stars inaccessible by interferometric observations.
The ultraviolet flux distribution for $\alpha$ Eri has been measured by the 
OAO-2 satellite 
\citep{code79}, giving us the effective temperature and luminosity of 
the star, as well as a SED for comparison with our models. 
All of these observations give us most of the information we need 
about the surface properties to model the star.

In the next section we present a more detailed description of the three
numerical components.
We discuss the surface results of the stellar evolution calculations and 
compare them to von Zeipel's law in \S 3. 
In \S 4 we discuss the effect of various parameters on the model atmospheres
as well as what was adopted and why.
The synthetic SEDs produced are discussed
in \S 5, and our conclusions are presented in \S 6.

\section{The Codes}

We wish to examine the characteristics of two rotating models with very 
different structure but with nearly the same horizontal average 
effective temperature and luminosity.  Three codes are required to generate
simulated SEDs for these models: the 2D stellar evolution code to provide the
variation in surface values, the stellar atmosphere code to generate the single
temperature SEDs and the code to integrate individual SEDs over the stellar 
surface.  
The interior and surface results alone are used to determine constraints on 
the applicability of von Zeipel's law.  We describe the properties of the three
codes in turn.

\subsection{2-D Stellar Evolution:  {\tt ROTORC}}
\label{rotorc}

The surface variation of T$_{eff}$ and g$_{eff}$ used to determine the size of
the required grid of model atmospheres is taken from stellar evolution
sequences computed with the 2.5D finite difference stellar evolution and 
hydrodynamics
code, {\tt ROTORC} \citep{bob90,bob95,bob98}. The half dimension means that
there is an equation for the azimuthal component of the momentum, but
that the model is constrained to be azimuthally symmetric. Thus, the equations
to be solved are the time dependent conservation laws for mass, three 
components of momentum, energy and hydrogen abundance, as well as Poisson's
equation.  The independent
variables of the code are the fractional surface equatorial radius and
the spherical polar coordinate, $\theta$. The primary dependent variables are
the density, temperature, three velocity components, hydrogen mass fraction and
gravitational potential.  The calculations are performed in the inertial
frame, so
the azimuthal momentum equation in principle allows the rotational
velocity profile inside the star to be evolved without forcing it to be
uniform or even conservative. The code was developed in this way so it can be 
used to
perform implicit hydrodynamic simulations, albeit in 2D,  to determine
any hydrodynamic or secular redistribution of angular momentum. 
All models calculated here included core overshooting of 0.38 of the pressure
e-folding distance at the convective core boundary,
based on the 2D hydrodynamic simulations of \citet{bob00,bob01}. The
radiative opacities and equation of state are calculated from the OPAL
tables \citep{opal}. 

The Zero-Age Main Sequence (ZAMS) models are calculated by specifying the 
rotational velocities as a 
function of the fractional radius and spherical polar coordinate $\theta$
and then solving for the gravitational potential, density and temperature
distributions inside the star.  Stellar evolution is performed in the usual
way with two exceptions.  Firstly, there are three components to the momentum 
equation and secondly, the 
time dependent composition equations are solved simultaneously in the implicit
Henyey solution instead of explicitly outside it.
The evolution sequences presented  here have been calculated with local 
conservation of angular
momentum throughout the evolution. Until the very end of the main
sequence, the angular momentum distributions calculated by forcing the
convective core to rotate uniformly and forcing angular
momentum to be conserved locally are nearly the same because the structure 
of the convective core
does not change much during this early evolution.  We have imposed equatorial
symmetry for better angular resolution with our angular zoning.

The features of primary interest here relate to the determination of the
physical conditions at the surface of the stellar model. The stellar surface
has traditionally been treated somewhat cavalierly in stellar structure
and evolution codes, and {\tt ROTORC} is no exception.  For example, the 
radiative flux in the energy conservation equation is calculated using the
2D diffusion equation, even in the optically thin regions.  At the 
surface, we set the flux to be equal to 2$\sigma$T$_{surf}^4$, where T$_{surf}$
is the temperature in the radial zone that defines the surface at each 
latitude (see \S 3 for discussion).  In addition to the usual crude surface 
conditions, spherical 
diversion makes the angular zoning near the surface quite coarse, so the 
surface is poorly resolved in the angular coordinate, although the surface 
outline is not unreasonable.  Some criterion must be 
stipulated to define the surface as a function of angle, and we have chosen
to make the surface an approximate equipotential.  The surface is an 
equipotential for conservative
rotation laws, and is not unreasonable in general except when the 
evolutionary phase is so rapid that the stellar surface might not be able
to adjust to the equipotential configuration.  We take the total potential,
$\Psi$, to be given in terms of the gravitational potential, $\Phi$, and the 
rotational `potential' by
\begin{equation}
\Psi = \Phi - \frac{1}{2} v^2_{\phi}
\label{eq:pot}
\end{equation}
The gravitational potential is determined by evaluation of the
gravitational potential exterior to the stellar surface as a surface
boundary condition and solving Poisson's equation throughout the entire
2D mesh.  Our models assume the second term of Eqn.~\ref{eq:pot} is a 
potential, but only at the model surface. This assumption will not necessarily
be true for arbitrary rotation laws.  The evaluation of the surface
gravitational potential is included in the Jacobian generated by the 
Henyey perturbation technique as the appropriately weighted integral over the 
mass distribution. For 
oblate spheroids, the surface value of
$\Psi$ is chosen as that value at the equator. The fractional radius of the
surface at each angle is chosen as the fractional radius of the radial
zone closest to the surface value of $\Psi$. Because the location of the
surface is slightly quantized by the radial zoning in this way (i.e., the
surface radius assumed is not quite the radius at the location of the
desired value of the total potential), the value of the effective
temperature may be slightly in error.  The error may amount to $\sim$ 200K 
in rapidly rotating models.  One result of this approach is 
that {\tt ROTORC} deals only with the potential and calculates its derivatives
as needed in the radial and latitudinal directions and therefore we need to
solve for the direction of the surface normal at each angle in our 
subsequent calculations.

A key result of these 2D simulations is that we obtain values of the 
effective temperature, surface radius, gravitational potential, and `total' 
potential (as defined by Eqn. \ref{eq:pot}) as a 
function of colatitude. This is something that 1D evolution codes, even 
those which include some of the effects of rotation (e.g., \citet{pins89,mnm}), do not provide except under very special 
circumstances. These quantities can be provided by the self consistent 
field method \citep{ost68} calculations of \citet{jack05}, but their code is 
currently only a stellar structure, not 
a stellar evolution, code, and their models to date are only ZAMS models.
This variation in surface quantities is required to generate the model 
atmospheres described in \S \ref{phoenix}.


\subsection{Synthetic Atmospheres: {\tt PHOENIX}}
\label{phoenix}
To generate our model atmospheres, we use the non-LTE atmosphere code 
{\tt PHOENIX}.
{\tt PHOENIX} makes use of a fast and accurate Operator Splitting/Accelerated Lambda Iteration
(OS/ALI) scheme to solve self-consistently the radiative
transfer equation and the NLTE statistical equilibrium (SE) rate equations for many species and overlapping transitions
\citep{hauschildt_b99} in a stellar atmosphere.
\citet{short_hb99} have greatly
increased the number of species and ionization stages treated in
SE by {\tt PHOENIX}. 
At least the lowest two stages of 24 elements,
including the lowest six ionization stages of the 20 most
important elements, including Fe and four other \ion{Fe}{0}-group elements, are now treated in NLTE.  
\citet{short_hb99} contains details
of the sources of atomic data and the formulae for various atomic processes.

\clearpage

\begin{deluxetable}{lrrrrrrrr}
\footnotesize
\tablecaption{Species treated in Non-Local Thermodynamic Equilibrium (NLTE)
in the NLTE$_{\rm Light}$ and NLTE$_{\rm Fe}$ models.
Each ionization stage is labeled with the number of energy levels and bound-bound
transitions included in the statistical equilibrium rate equations.  Note that
this table shows only a sub-set of the total number of species that are 
currently treatable in statistical equilibrium by {\sc PHOENIX. }
\label{table:t1d}
}
\tablecolumns{9}
\tablewidth{0pt}
\tablehead{
\colhead{Element} & \multicolumn{1}{c}{Model} & \multicolumn{4}{c}{Ionization Stage} \\
\colhead{} & \colhead{} & \colhead{\ion{}{1}} & \colhead{\ion{}{2}} & \colhead{\ion{}{3}} & \colhead{\ion{}{4}} }
\startdata
H   & NLTE$_{\rm Light}$, NLTE$_{\rm Fe}$     &  80/3160 &\nodata &\nodata &\nodata  \\
He  & NLTE$_{\rm Light}$, NLTE$_{\rm Fe}$     &  19/37 & 10/45  &\nodata &\nodata  \\
Li  & NLTE$_{\rm Light}$, NLTE$_{\rm Fe}$     &  57/333 & 55/124 &\nodata &\nodata  \\
C   & NLTE$_{\rm Light}$, NLTE$_{\rm Fe}$     &  228/1387 & 85/336 & 79/365 &\nodata \\
N   & NLTE$_{\rm Light}$, NLTE$_{\rm Fe}$     &  252/2313 & 152/1110 & 87/266 &\nodata \\
O   & NLTE$_{\rm Light}$, NLTE$_{\rm Fe}$     &  36/66 & 171/1304 & 137/765 &\nodata \\
Ne  & NLTE$_{\rm Light}$, NLTE$_{\rm Fe}$     &  26/37 &\nodata &\nodata &\nodata \\
Na  & NLTE$_{\rm Light}$, NLTE$_{\rm Fe}$     &  53/142 & 35/171 & \nodata  &\nodata \\
Mg  & NLTE$_{\rm Light}$, NLTE$_{\rm Fe}$     &  273/835 & 72/340 & 91/656 &\nodata \\
Al  & NLTE$_{\rm Light}$, NLTE$_{\rm Fe}$     &  111/250 & 188/1674 & 58/297 & 31/142 \\
Si  & NLTE$_{\rm Light}$, NLTE$_{\rm Fe}$     &  329/1871 & 93/436 & 155/1027 & 52/292 \\
P   & NLTE$_{\rm Light}$, NLTE$_{\rm Fe}$     &  229/903 & 89/760 & 51/145 & 50/174 \\
S   & NLTE$_{\rm Light}$, NLTE$_{\rm Fe}$     &  146/439 & 84/444 & 41/170 & 28/50 \\
K   & NLTE$_{\rm Light}$, NLTE$_{\rm Fe}$     &  73/210 &  22/66 & 38/178 &\nodata\\
Ca  & NLTE$_{\rm Light}$, NLTE$_{\rm Fe}$     &  194/1029 & 87/455 & 150/1661 &\nodata \\
Fe  & NLTE$_{\rm Fe}$     &  494/6903 & 617/13675 & 566/9721 & 243/2592 \\
\enddata
\end{deluxetable}

\clearpage

\begin{table}
\caption{Levels of modeling realism.
\label{table:deg}
}
\begin{tabular}{lll}
\tableline
Degree of NLTE          & Model designation\\
\tableline
None                    & {LTE} \\
Light metals only (H-Ca)& {NLTE$_{\rm Light}$} \\
Light metals \& \ion{Fe}{0}   & {NLTE$_{\rm Fe}$} \\
\tableline
\end{tabular}
\end{table}

\clearpage

Table \ref{table:t1d} shows which species have been treated in NLTE in the modeling presented here, and how many E levels and $b-b$ (bound-bound) transitions are included in 
SE for each species.  E is defined as the energy of the state with respect to
the ground state of that ionization stage.  
Table \ref{table:deg} explains which elements are included in the degrees of realism modeled.  
For the species treated in NLTE, we use the factor $gf$, where $g$ is the
statistical weight of the lower level and $f$ is the oscillator strength of the
transition.  This factor is read in from the line lists used by {\tt PHOENIX}.
Only levels connected by transitions of $\log gf$ value
greater than -3 (designated primary transitions) are included directly in the SE rate equations.  
All other transitions of that species (designated secondary transitions) are calculated
with occupation numbers set equal to the Boltzmann distribution value with excitation 
temperature equal to the local kinetic temperature, multiplied by the ground state 
NLTE departure coefficient for the next higher ionization stage. 
We have only included in the NLTE treatment those ionization stages that are non-negligibly 
populated at some depth in the star's atmosphere.  As a result, we only include the
first one to four ionization stages for most elements.  Additionally, tens of millions of 
transitions are included with the approximate
treatment of LTE.

NLTE effects can depend sensitively on the adopted values of atomic parameters that affect the 
rate of collisional and radiative processes.  Atomic data for the energy levels and b-b transitions have been taken from \citet{kurucz94b} and \citet{kurucz_b95}.
We have used the resonance-averaged Opacity Project \citep{seaton_ymp94} data of
\citet{bautista_rp98} for the ground-state photo-ionization cross sections of \ion{Li}{1}-\ion{}{2}, \ion{C}{1}-\ion{}{4}, \ion{N}{1}-\ion{}{6}, \ion{O}{1}-\ion{}{6}, \ion{Ne}{1}, \ion{Na}{1}-\ion{}{6}, \ion{Al}{1}-\ion{}{6}, \ion{Si}{1}-\ion{}{6}, \ion{S}{1}-\ion{}{6}, \ion{Ca}{1}-\ion{}{7}, and \ion{Fe}{1}-\ion{}{6}.  
For the ground states of all stages of P and Ti
and for the excited states of all species, we have used the cross sectional data previously incorporated into
{\tt PHOENIX}, which are from \citet{reilman_m79} or those compiled by \citet{mathisen84}.  We account for
coupling among {\it all} bound levels by electronic collisions using cross sections calculated with the 
formula of \citet{allen73}. 
We do not use the formula of \citet{vanreg62} for pairs of levels that are connected by a permitted
radiative transition because we have found that doing so leads to rates for
transitions within one species that are very discrepant with each other,
and this leads to spurious results.  
The cross sections of ionizing collisions with electrons are calculated with the formula of \citet{drawin61}.  

For our models, we calculated intensity grids covering the region 1000 $\mbox{\AA}$
to 4000 $\mbox{\AA}$ with $\Delta \lambda$ = 0.02 $\mbox{\AA}$, giving a resolution of
R = $\lambda / \Delta \lambda$ = 150000.  

\subsection{The Atmospheric Integrator}
\label{integ}

Once we have the individual model atmospheres, we must produce 
an integrated flux spectrum for a model with non-uniform surface parameters.
The intensity grid results from the NLTE model atmospheres are then convolved 
with the instrumental profile of the OAO-2 satellite \citep{code79}.  Each 
spectrometer has a
response function covering about 40 $\mbox{\AA}$, so this convolution smoothes
over the individual lines.  
For this reason, we bin the convolved intensity files to 10 $\mbox{\AA}$ 
resolution.  At this resolution, the individual lines are not visible, 
and the bins are large enough that the effects of the Doppler shift are not 
significant.

The input to this code comes from stellar evolution models generated by 
{\tt ROTORC}\@.  We have evolved two specific models to match the approximate
observed temperature 
and luminosity of Achernar, but with differing degrees of oblateness.  From 
these
models, we were able to generate effective temperatures and gravities as a 
function of latitude.  These values determined the range of the atmospheric
grid produced by {\tt PHOENIX}\@.  We used a grid with temperature range of 
11000 to 25000K with 2000K spacing and a range in log {\it g} of 2.3 to 3.7, 
with spacing of 0.2.  This range of temperatures is required to produce
synthetic spectra of our models of Achernar.

For each wavelength, we wish to evaluate the integral
\begin{equation}
F_{\lambda} = \int_{\theta} \int_{\phi}  \frac{I_{\lambda}(\xi (\theta,\phi))}{d^2} dA_{proj}
\end{equation}
where $\theta$ is the colatitudinal coordinate, $\phi$ is the longitudinal 
coordinate, $\xi(\theta,\phi)$ is the angle between the local surface normal 
and the line of site to the observer, d is the distance to the object, 
dA$_{proj}$ is the projected surface area element as seen from the direction
of the observer,
I$_{\lambda}$ is the intensity at a given wavelength and F$_{\lambda}$ 
is the flux at a given wavelength.

To do this integration, the surface parameters are read in from the output of 
{\tt ROTORC}\@.  
The surface of the star is then divided into a mesh, typically 200 $\theta$
zones and 400 $\phi$ zones.  
For each zone, the effective temperature and surface gravity are determined 
from the {\tt ROTORC} model.  
The atmospheric integration code reads in the appropriate intensities from a 
grid of 
models in T and log {\it g} produced by {\tt PHOENIX} and performs linear
interpolation over log {\it T} and log {\it g} to determine the intensity 
produced by each grid zone.  

Next, the angle between the local surface normal and the line of sight to the
observer ($\xi$) is determined as follows.
The model is axisymmetric, so the observer can be assumed to be directly above
the prime meridian ($\phi$ = 0) of the star with no 
loss of generality.
This gives the vector coordinates from the prime meridian towards the observer
of $\delta$x = sin{\it i},  $\delta$y = 0 and $\delta$z = cos{\it i}.

\clearpage
\begin{figure}
\center
\plotone{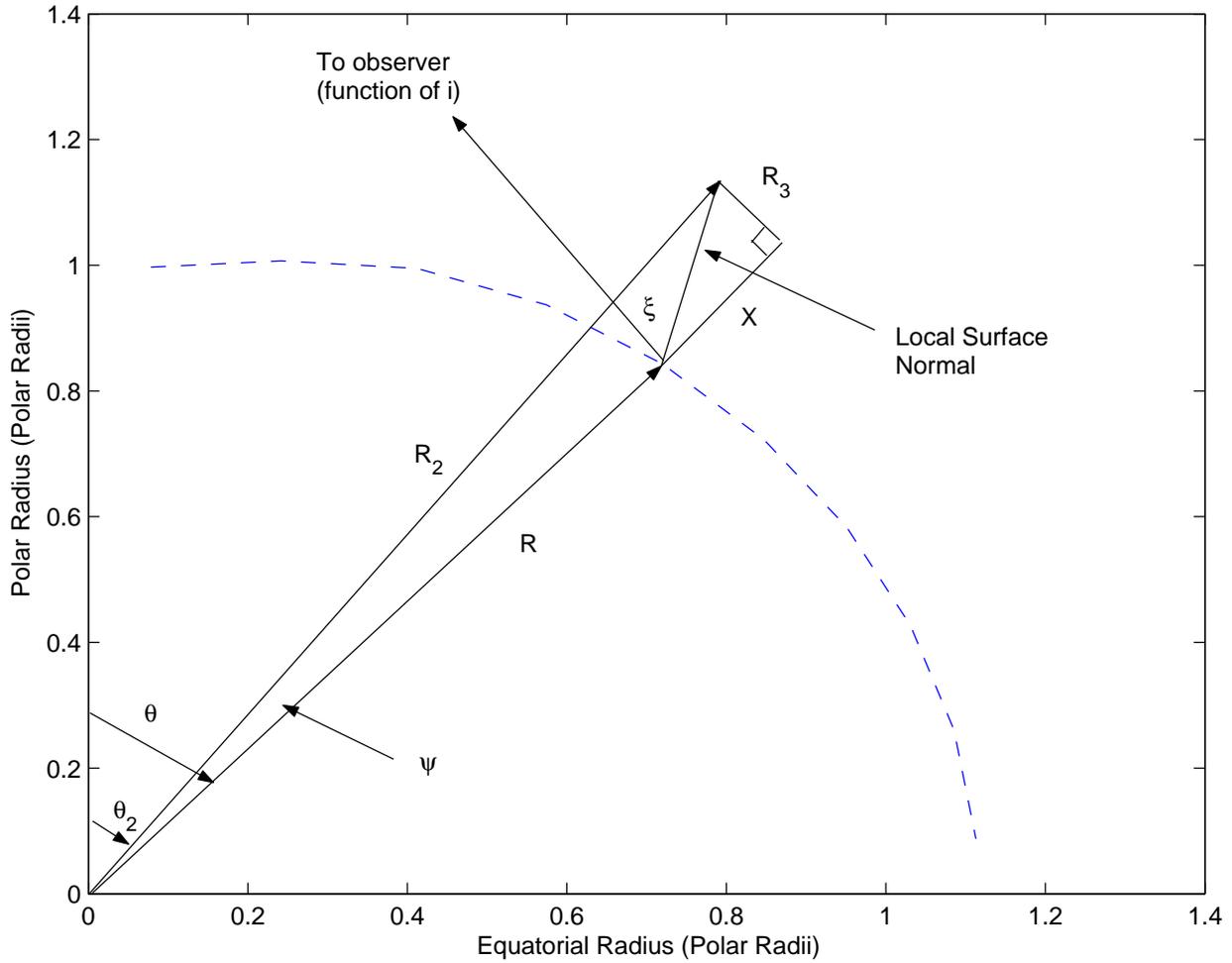}
\caption{Schematic diagram of the geometry used to determine the vector 
perpendicular to the surface.  The distance from the model center to the 
surface at the location of interest is R.  This vector is extended by an 
arbitrary length X.  R$_3$ is perpendicular to X and is bounded by the vector 
X and the vector perpendicular to the surface.  The vector R$_2$ runs from 
the model center to the intersection of R$_3$ with the surface normal.  The 
direction of the surface normal is given by the difference between vectors 
R$_2$ and R.  The dashed line shows a sample surface geometry.}
\label{fig:geo}
\end{figure}
\clearpage

To find the surface normal, we refer to Fig.~\ref{fig:geo}.  
Starting with the radius at a given point on the 
model surface, R, we extend this vector an arbitrary distance X.  
Next, we extend the surface normal until it meets a vector R$_3$, 
perpendicular to X\@.
The intersection of R$_3$ and the surface normal occurs a distance R$_2$ from 
the center of the model.  
The difference between the two vectors R$_2$ and R is in the direction
normal to the surface.
If R has a polar angle $\theta$ and R$_2$ has a polar angle $\theta$$_2$, then 
\begin{equation}
\theta_2 = \theta - \psi
\end{equation}
where $\psi$ is positive in the northern hemisphere and negative in the
southern hemisphere as a result of the oblate shape of the model.  
The angle, $\Lambda$, between the radial vector (X) and the surface normal 
can be approximated by 
\begin{equation}
tan(\Lambda) \approx \frac{\delta R(\theta)}{R\delta \theta}
\label{eqn:tanl}
\end{equation}
where $\delta$R is the the variation of the surface radius over an angle 
$\delta\theta$.  

The vector normal to the surface can be defined by the spherical coordinates 
(R,$\theta$,$\phi$) and (R$_2$,$\theta$$_2$,$\phi$), of which R, $\theta$,$\phi$ 
and $\Lambda$
are known and X is assumed.
From these, we can calculate
\begin{equation}
R_3 = X tan \Lambda
\end{equation}
\begin{equation}
R_2^2 = (R+X)^2 + R_3^2
\end{equation}
\begin{equation}
\psi = arcsin(\frac{R_3}{R_2})
\label{eqn:psi}
\end{equation}
Given these quantities we can perform the dot product of the vector R$_2$
-R with the line of sign vector to find cos$\xi$.
There are other ways by which the normal could be calculated.  One of these
would be to use the vector sum of the gravitational and centrifugal forces,
as this sum is normal to the equipotential surface.  This would then be 
interpolated between the centers of the angular zones. We decided to use the
equipotential as defined by Eqn\@. \ref{eq:pot} because this is what the
2D code uses to determine the surface location.  Given the collection of
approximations made in the calculation, we do not regard the uncertainties in
this aspect of the calculation as sufficiently significant to investigate
different methods.

Equations \ref{eqn:tanl} to \ref{eqn:psi} allow us to calculate the direction 
cosine between the surface normal 
and the vector pointing towards the observer, $\xi$.
By definition the surface is not visible to the observer if 
cos$\xi$ $<$ 0.  

Once cos$\xi$ has been determined, an interpolation over the angles in the intensity files is performed.  
This gives the contribution to the total flux per wavelength from each 
grid zone. 
This total flux is then weighted according to the projected surface area for 
each mesh zone
\begin{equation}
dA_{proj} = R(\theta)^2sin\theta cos\xi\sqrt{1+\Big(\frac{dR}{R d\theta}\Big)^2} d\theta d\phi
\end{equation}
The process is then repeated for each wavelength.  
For these models, we calculated the flux for every 10 $\mbox{\AA}$ for wavelengths
covered by the OAO-2 spectrometers, 1160 to 3600 $\mbox{\AA}$\@.
This allows for direct comparison with the UV spectrum taken by
the OAO-2 satellite.  
However, this is not a limitation on the code, and
any wavelength range or spacing could be used.   

To ensure that the integrator worked correctly, we compared the final 
\emph{flux} spectrum for a uniform sphere produced by {\tt PHOENIX} and by our
atmospheric integrator.  
{\tt PHOENIX} performs the integration by adding up the contributions of a 
series of concentric annuli \citep[pp 11-12]{mihalas}, while our integrator 
uses a mesh in $\theta$ 
and $\phi$.  This method produces a finer mesh in the polar regions than 
near the equator.  The two flux spectra are very similar overall, although 
there are some slight variations.  These variations are thought to be due
to slight numerical differences in the methods of integration.  Another 
difference between these two models is the order of operations.  In our model,
we convolve the 0.02 $\mbox{\AA}$ spaced intensity grid and then integrate the 
product.  In the {\tt PHOENIX} model, the SED is calculated at 0.02 $\mbox{\AA}$
and then convolved.  We checked that these two operations commute by 
integrating a small section of the intensity grid at a resolution of 0.02 
$\mbox{\AA}$ and then comparing
it to the {\tt PHOENIX} flux spectrum.
The two unconvolved spectra differ by about 0.8\% over a region spanning 150 
$\mbox{\AA}$\@.

We also tested the spacing of our grid.  Initially, our models were spaced
at intervals of 0.2 in log{\it g} and 2000 K in temperature.  In general,
we found the difference between successive gravities was very small, so
we concluded the resolution in log{\it g} was sufficient.  
We used our integrator to produce a SED for a model at 12000 K
based on intensity files at 11000 and 13000 K\@.  Next, we compared this to a 
SED based on the the intensity files at 12000 K\@.  To 
estimate
how accurate the interpolation was, we took the ratio of the two 12000 K 
models.  The fourth root of this ratio gave us an estimate of the ratio of
the temperatures corresponding to these fluxes.  On average, the ratio
calculated was 0.98, corresponding to a 2\% error in the temperature.  We 
concluded that this amount of error was acceptable, and hence our temperature
spacing of 2000 K was sufficient.

\section{Comparison of Evolutionary Surface Results with von Zeipel's Law}
\label{sec:roche}

A general way of exploring the effective temperature variation on the 
surface of a rotating star under specific assumptions was outlined by 
\citet{vonz}. If the 
centrifugal acceleration is conservative, it can be written as the gradient 
of a potential. This means that the gradient of the pressure is given by 
the density times the gradient of the sum of this potential and the 
gravitational potential. Thus, the pressure is 
constant on the equipotential surface, and the density must be as well. If 
the equation of state is a function of the density, temperature, and 
composition, then the temperature will also be constant on the 
equipotential surface if the composition is uniform. If the energy transport 
is by radiation and the diffusion approximation may be used for the radiative 
flux, then the energy flow must be perpendicular to the surfaces of constant 
temperature, i.e., perpendicular to the equipotential surfaces. This flux 
can be written in terms of the gradient of the total potential, which is 
just the effective gravity. At the surface this flux is proportional to the 
fourth power of the effective temperature, so we have
\begin{equation}
T_{eff} \propto g_{eff}^{0.25}
\label{eq:vonz}
\end{equation}

Here we wish to compare the results of our evolution calculation surfaces 
with those based on this simple model. There are several possible sources of 
disagreement. 
One feature that the simple model 
fails to treat is the coupling of the effective temperature to the surface 
temperature structure in any way. This surface temperature structure will 
not alter the temperature structure of the model much, except near the surface,
but it could play a role in situations near critical rotation, where the 
von Zeipel model predicts that equatorial 
effective temperature vanish as the effective potential goes to zero, or in 
other situations in which there is significant variation in effective 
temperature between the pole and equator.

To make this comparison we will examine four models from two evolutionary 
sequences. Two of the models are ZAMS models, one for a uniformly rotating 
model near critical rotation, and the other for a model with significant 
differential rotation. 
We use the parameter $\eta$
\begin{equation}
\eta = \frac{\Omega^2 R_e^3(\Omega)}{GM},
\end{equation}
the ratio of the centrifugal and gravitational forces at 
the equator.  At critical rotation, this parameter should have the value 1.
The uniform rotation ZAMS models has a surface equatorial velocity of 495 km
s$^{-1}$ and $\eta$ =  0.86.
The ratio of polar to equatorial radius is 0.70.  The differentially rotating
model had a surface equatorial velocity of 410km s$^{-1}$ on the ZAMS, which
gives $\eta$ = 0.56.  The ratio of the polar to equatorial radius is 0.78.  
Each of these ZAMS models is then evolved through core hydrogen burning and 
a model with the average luminosity and effective temperature close to 
those observed for Achernar chosen. These two evolved models will be 
compared with the observed SED of Achernar as well as each other.
Because of the effects of rotation on the surface properties, different 
rotation laws require different masses to reproduce the same average
surface temperature and total luminosity.  As the rotation increases, the
model moves down and to the right in the HR digram when plotted using its 
``average'' surface quantities.  To compensate 
for this effect, the mass must be increased to raise the average luminosity.
Increasing the internal angular 
momentum for a given surface velocity has the same effect.
Our models are not as oblate as those described in \citet{jack04} because
we are not yet able to model such extreme angular momentum distributions.  
The 2D code currently expects the equator to have the largest radius, and 
for very high angular momentum models, such as the distributions described 
in \citet{jack04}, this is not the case.

These evolutionary sequences locally conserve angular 
momentum.  This is different from the usual assumption of forcing uniform 
rotation in the convective core.  However, the two approaches give very nearly
the same result until the very end of core hydrogen burning is reached, because
the density structure in the core does not change significantly until that 
stage is reached.  It should be noted that the evolved models will 
not have conservative rotation laws, although the departures from a law which 
depends only on distance from the rotation axis are slight.


The surface effective gravity is calculated in a relatively straight forward
way.  From the surface shape, we can determine the angle between the radial
and the normal to the surface.  Because we know both the radial and latitudinal
variation of both the gravitational and centrifugal forces, the effective
gravity follows.

\begin{figure}
\center
\plotone{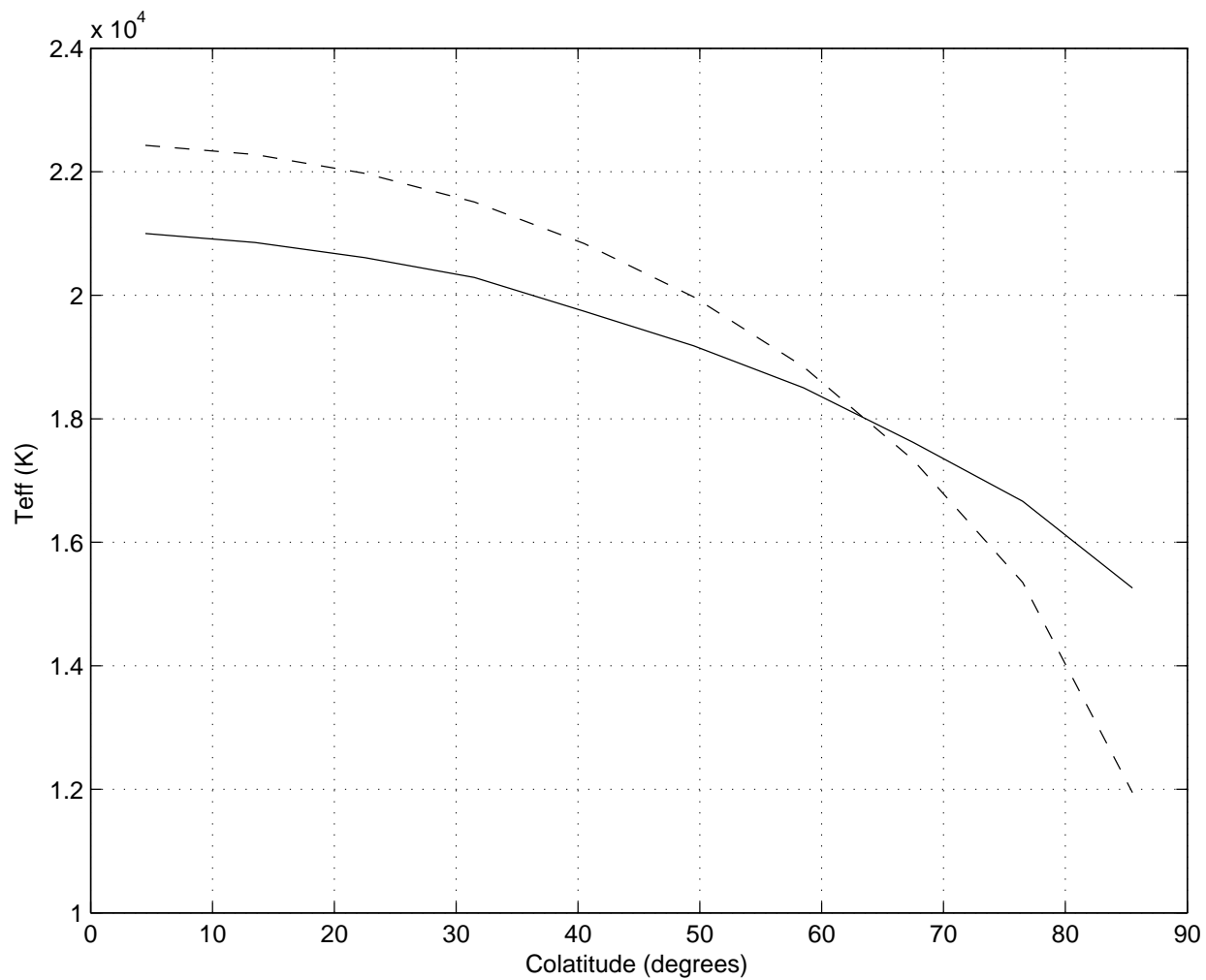}
\caption{Comparison of the effective temperature variation for the 6.5 
M$_{\odot}$ ZAMS model rotating as a solid body near critical rotation (solid)
and from von Zeipel's law
(dash). von Zeipel's law is calibrated so that the total flux radiated from
the surface is the same as that for the {\tt ROTORC} model. The von Zeipel 
equatorial flux may be too low because the effective temperature is not 
coupled to the temperature near the surface of the model.}
\label{fig:surf}
\end{figure}

First we examine the uniformly rotating case. The variation of the {\tt ROTORC}
gravitational potential on a sphere with radius equal to the surface 
equatorial radius is 
quite small, only about 0.1\% between the pole and equator. 
This is not surprising because even near critical uniform rotation the inner
regions do not rotate sufficiently fast to produce any significant 
horizontal variation in the mass distribution which would show up in the 
surface gravitational potential. A comparison of the effective temperatures 
from {\tt ROTORC} and from von Zeipel's law are shown as a function of 
colatitude Fig. \ref{fig:surf}. The effective 
temperatures calculated from von Zeipel's law show a greater range than do 
those from the 2D model, being both larger at the pole and smaller at the 
equator.  At least part of this arises from the conditions at the equator; the 
{\tt ROTORC} surface flux there is more than four times greater than the von 
Zeipel
model flux.  This must be compensated for somewhere, as the luminosity emitted 
by the two models is required to be the same.  As a result, the von Zeipel 
polar temperature and flux are higher than the {\tt ROTORC} values.

\clearpage
\clearpage

We believe the source of the disagreement between our calculation and the von 
Zeipel model can be found in a contradiction in the von Zeipel model.  The
von Zeipel model argues that the temperature is constant on an equipotential
surface, but the effective temperature varies significantly from the pole to
the equator.  This can be true only if the effective temperature is completely
independent of the surface temperature structure of the star.  Our 
models require the temperature in the last zone to be given by the surface
($\tau$ = 0) temperature of a simple grey atmosphere:
\begin{equation}
T(\tau)^4 = \frac{3}{4}T_{eff}^4(\tau+\frac{2}{3})
\end{equation}
so the surface flux is just 2$\sigma$T($\tau$=0)$^4$.

To determine if this could be the source of the differences in 
Fig.~\ref{fig:surf}, we integrated model envelopes (in the stellar structure
sense) from the surface inward.  A common approach is to assume values of
L, T$_{eff}$, M and the composition and start with a very small density,
$\rho$($\tau$ = 0), at r = R.  Although a number of variations are possible,
most do not matter unless the envelope is highly extended.  Our integration
scheme is modeled on a code \citet{pacz69} developed to
produce outer boundary conditions for his stellar evolution codes, but the 
code has been updated to include the same physics as {\tt ROTORC}.  We 
modified the envelope integrator to decouple the effective temperature and
the surface temperature by treating the surface temperature as a free 
parameter.  We then compared the density, temperature and pressure 
distributions of model envelopes for a given effective temperature but with 
a range of surface temperatures.

As one would expect, the differences decrease with depth into the model.  
However, the differences only drop to about one percent at temperatures
of about 6$\times$10$^5$ K, which is about the variation we see along
equipotential surfaces at these temperatures.  This is not an ironclad 
argument because we have not included the centrifugal force in these envelope
calculations, but it is suggestive.

The evolved model shows that these differences have been significantly reduced,
as the {\tt ROTORC} model is now much less oblate.  The ratio 
of the polar to 
equatorial radius is now 0.81 and the rotational surface equatorial velocity 
has dropped
to 278 km s$^{-1}$, with $\eta$ = 0.45.  Note that this surface equatorial 
velocity is close to the
value of v sin{\it i} for Achernar, but the {\tt ROTORC} model is much less
oblate.  The range of temperatures predicted by the von Zeipel model
is still slightly larger, but in no location is the temperature difference 
between the two models greater than 350 K\@.  The discrete zoning of the 
{\tt ROTORC} models produces about a 200 K difference if the surface is changed
by one radial zone, so this uncertainty already accounts for a sizable 
fraction of this temperature difference.  Note that because of local
conservation of angular momentum, the {\tt ROTORC} rotation law is no longer
conservative after the evolution, but the departures are small with no 
obvious consequences.

The situation for strongly differentially rotating stars raises
different issues. For the rotation law, we have followed the general form
of \citet{jack04}:
\begin{equation}
\Omega = \frac{\Omega_{\circ}}{1+\alpha\varpi^n}
\label{eqn:power}
\end{equation}

The exponent was chosen to be 1.4 and the constant $\Omega_{\circ}$ chosen 
so that the surface equatorial velocity of the ZAMS model was 430 km s$^{-1}$.
The coefficient ($\alpha$) of the distance from the rotation axis ($\varpi$) 
is 2.0 in units where the surface equatorial value of $\varpi$ is unity. The 
rotation law is conservative and the model has significant differential 
rotation.

\begin{figure}
\center
\plotone{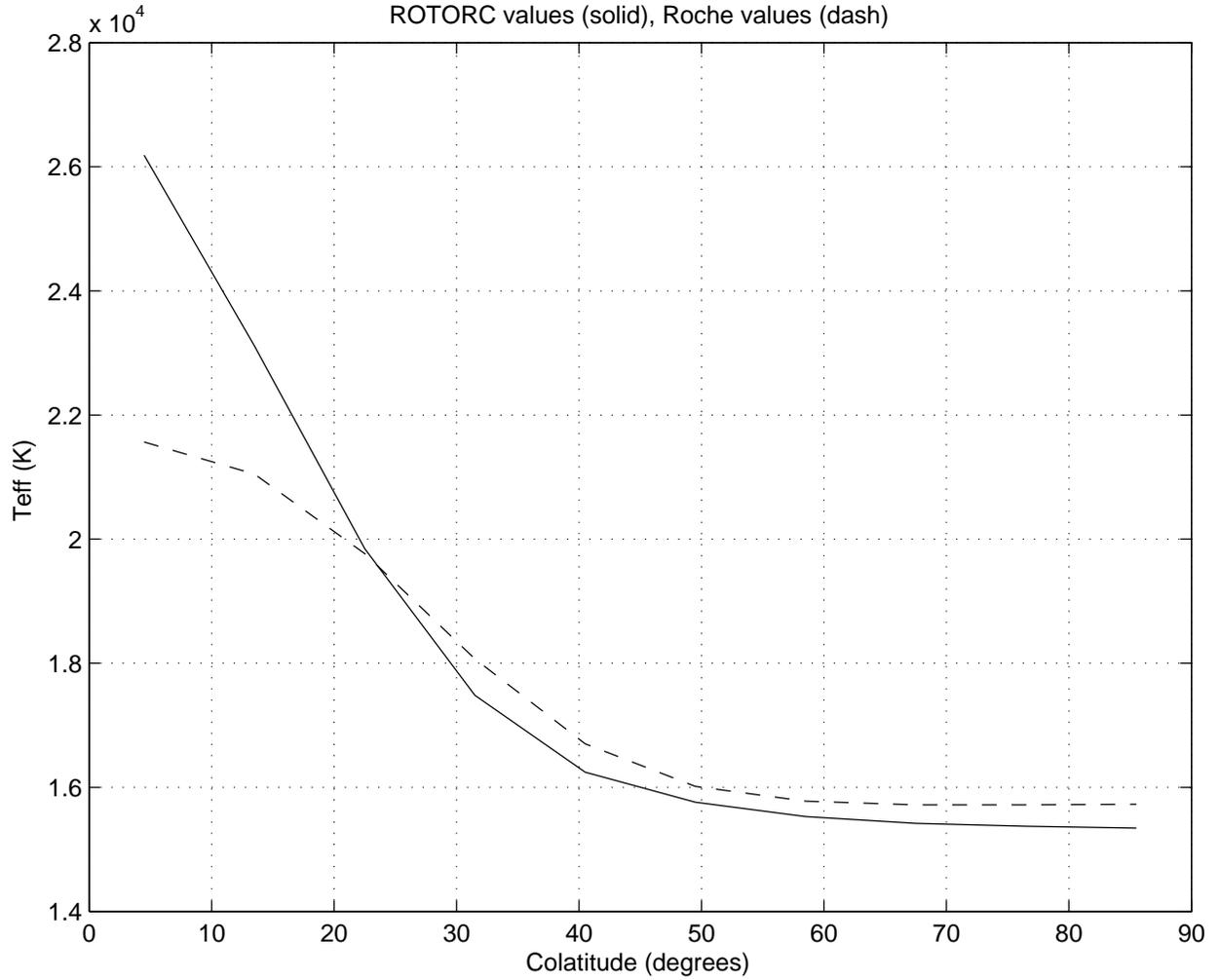}
\caption{Comparison of the effective temperatures for the 7 M$_{\odot}$ 
differentially rotating ZAMS model (solid) and von Zeipel's law (dash). The 
{\tt ROTORC} model clearly funnels more energy toward the high latitude regions
than von Zeipel's law would predict.}
\label{fig:surf2}
\end{figure}

Fig.~\ref{fig:surf2} compares the effective temperature as a function of
polar angle between the calculated {\tt ROTORC} model and von Zeipel's law.
The most significant difference is the higher temperature at the pole in
the {\tt ROTORC} model although both {\tt ROTORC} and von Zeipel's law show 
the same general shape in the latitudinal effective temperature dependence.

        A number of calculations were undertaken to determine the origin of 
the differences in Fig.~\ref{fig:surf2} and the
sensitivity of the computed results. Both the radial and angular zoning
resolution were increased appreciably, but there was no significant
effect on the surface temperature latitudinal dependence. The largest
effect of more refined zoning was the dropping of the variation of the 
density, pressure, and
temperature on equipotential surfaces (calculated after the fact). In the
relatively deep interior these variations were about 0.1\% with the
current zoning, and were reduced to about 0.03 \% when the angular
resolution was doubled. This amount is also about the departure of the 
numerical calculation of $\bigtriangledown^2(1/2(v_{\phi}^2))$ from $2\omega^2$
for the uniform rotation case.  The
radial zoning was already quite good and the changes of the variables on
equipotential surfaces was very slight. We also examined the calculation
of the radiative gradient, $\bigtriangledown_{rad}$, near the convective 
boundary by using
the actual computed pressure gradient instead of the usual stellar
structure expression based on spherical symmetry. The changes in the
uniformly rotating model were negligible because the rotation near the
convective core boundary is so low, but for this steep differential
rotation it did change the location of the convective core boundary
slightly, but again had no effect on the latitudinal variations of the
surface temperature. From these results and some other artificial
numerical exercises we conducted, we believe the differences between the 
surface
temperature latitudinal variations originates at or near the surface.

There are at least two surface possibilities to explain the differences. 
One is that, unlike the uniformly rotating case, the surface shape 
variations are fairly large close to the rotation axis, even though they 
must go to zero on the axis because of the axial symmetry. This makes the 
calculation of the normal to the surface somewhat uncertain and thus the 
contribution of the rotational potential in this direction is uncertain as 
well. Another is rooted in the same cause as for the uniform rotation case 
- the decoupling of the effective temperature from the surface temperature 
structure in von Zeipel's law.

One might wonder how the envelope could produce one situation in which von 
Zeipel's law shows greater variation than the 2D calculation in one case 
but less in the other. Part of the reason can be seen by comparing the 
polar and equatorial temperatures on equipotential surfaces as functions of 
depth into the model from the surface. In the uniform rotation case, the 
two temperatures started off at the surface values and progressively 
approached each other as the depth into the model increased. This is not 
true in the differential rotation case, where the significant difference in 
the opacity for the surface temperatures allows the two temperatures to 
cross on an equipotential surface on which the polar temperature is still 
optically thin because the opacity is appreciably lower. These two 
temperatures separate further as a function of depth although this 
eventually stops and  they gradually come together again at sufficient 
depth in the envelope. This temperature at which the polar and equatorial 
temperatures come together is effectively the same for uniform and 
differentially rotating model.

\clearpage
\clearpage

\section{Atmospheric Models}

Before we calculate our model atmospheres we must specify the composition.  
Several lines of evidence weakly point to using a composition which is 
approximately solar.   In this context a metal abundance of 0.02 is 
sufficiently 
close to the solar value of 0.0188.  Recent photospheric modeling points to 
a much lower iron abundance, based on rather uncertain oscillator 
strengths \citep{kostik96}.  We have chosen to use the solar abundance 
required to reproduce the helioseismology results for a 1 M$_{\odot}$ at the
solar age, around Z $\sim$ 0.018 \citep{antia05,bahc05}.
The observational evidence for a metallicity approximately solar includes the 
fact that Achernar is close to the sun (d = 44.1 pc) \citep{p97} and hence its 
metallicity is likely close to solar.  Another indication of the metallicity
of Achernar comes from a study by \citet{torres2000} which finds some evidence
for a lose association of pre-main sequence stars centered around ER Eri.  
Although this association consists primarily post-T Tauri stars, the age and
location of Achernar is consistent with a metallicity of Z = 0.02.  Finally,
many studies of galactic B stars indicate their average metallicity is close
to solar \citep{gehren85,brown86,lennon90}.  We have run comparisons of LTE
models with Z = 0.02 and Z = 0.04.  The higher metallicity models show more 
line blanketing, but the differences between the two SEDs are too small to 
have a preference of one metallicity over the other when compared to the
observed SED of Achernar.    Based on this admittedly weak evidence, we have
performed all our calculations with Z = 0.02.

We also calculated models making various assumptions about NLTE.  We have
compared models in which all energy levels are populated according to LTE,
models in which only the light elements are allowed to be in NLTE and models
in which the light elements and Fe are assumed to be in NLTE (see Table 
\ref{table:deg}).  The 
differences among the three resulting SEDs were sufficiently large and changed
the shape of the SED just enough that we felt that the models with both the
light elements and Fe in NLTE were needed.  The remaining discussion uses these
models.

\begin{figure}
\center
\plotone{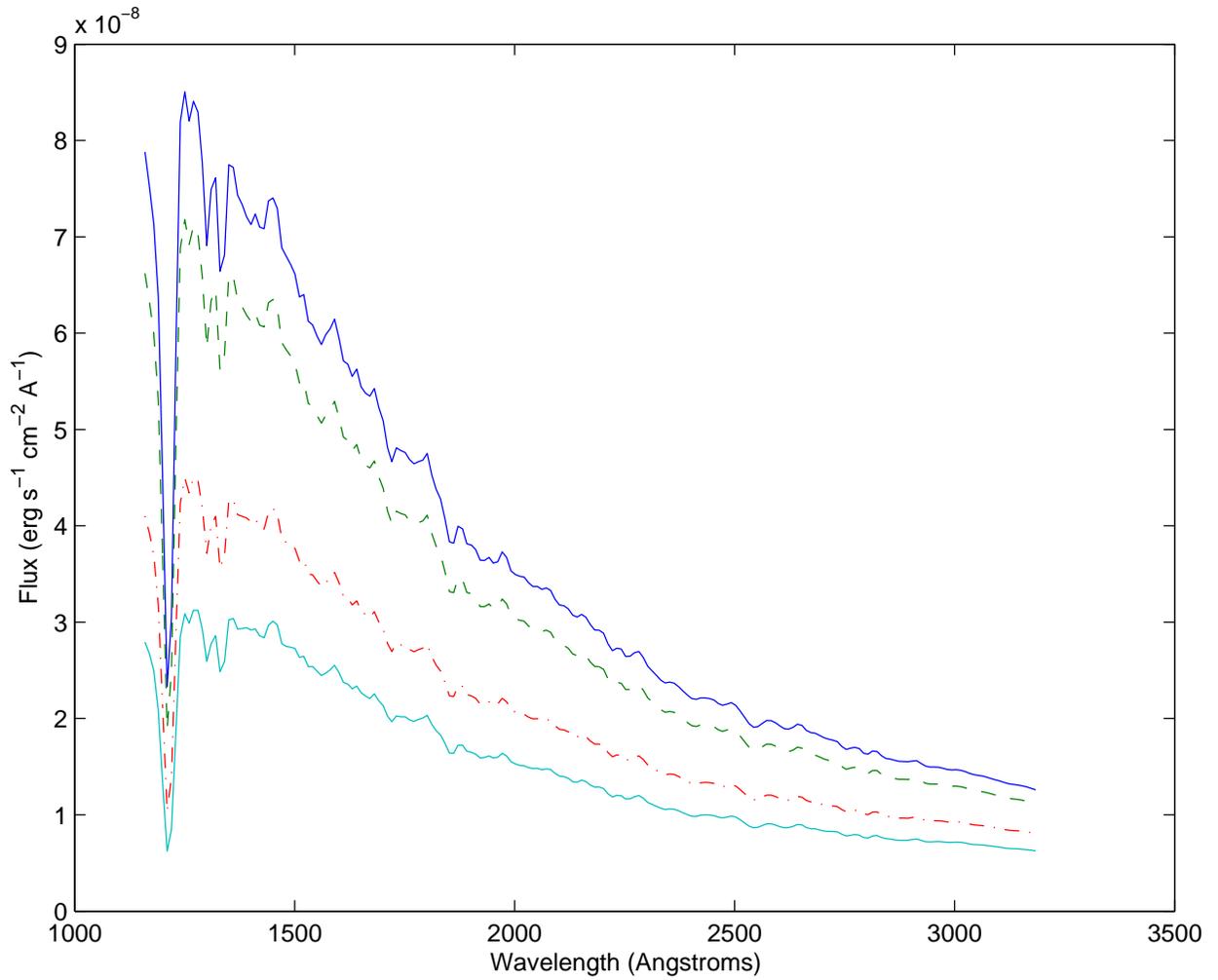}
\caption{Synthetic flux spectra for a 6.5 M$_{\odot}$ model at inclinations of 0$^{\circ}$
(top solid) 30$^{\circ}$ (dashed) 60$^{\circ}$ (dot dashed) and 90$^{\circ}$ (lower solid). } 
\label{fig:incline1}
\end{figure}

\begin{figure}
\center
\plotone{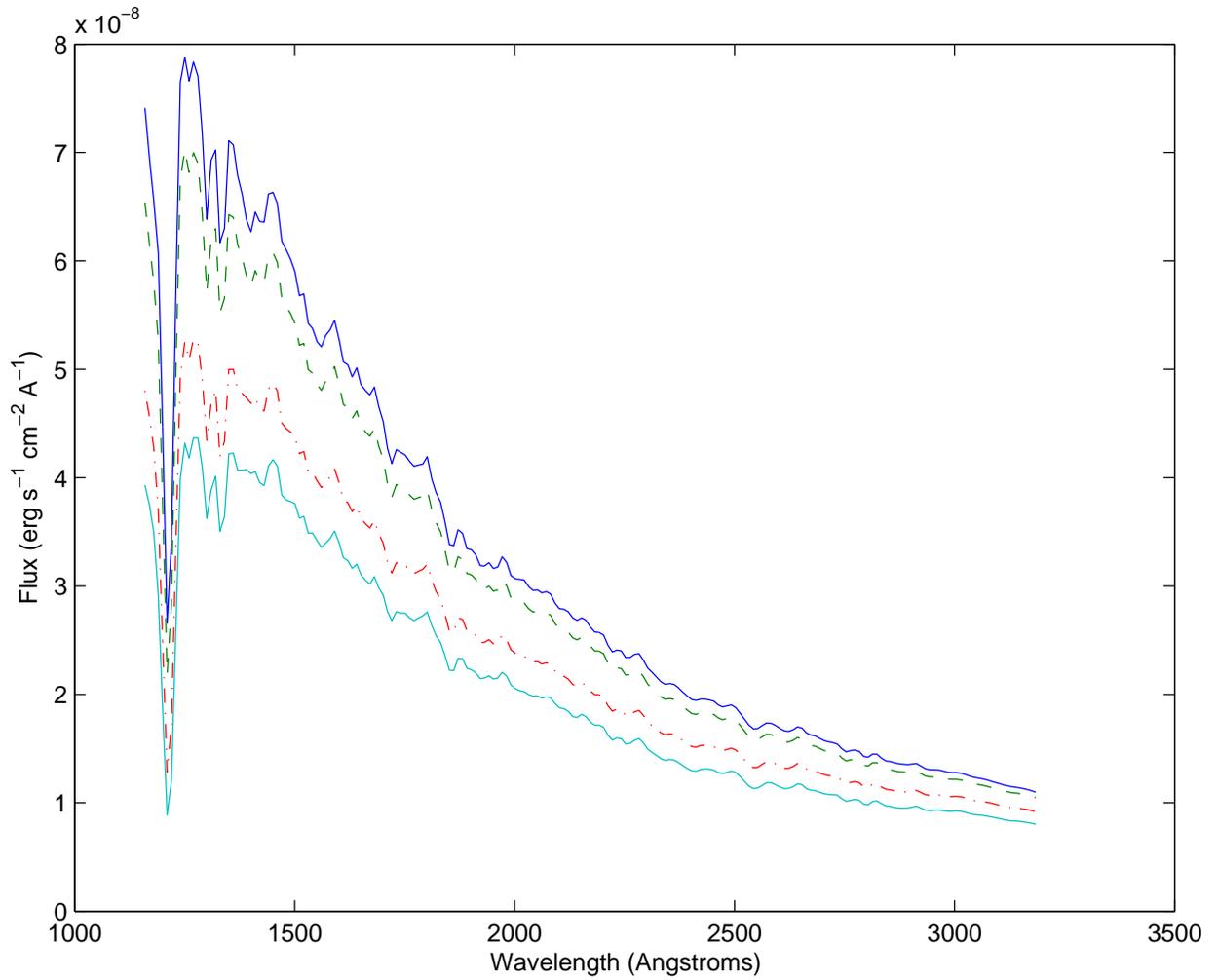}
\caption{Synthetic flux spectra for a 7.0 M$_{\odot}$ model at inclinations of  0$^{\circ}$
(top solid) 30$^{\circ}$ (dashed) 60$^{\circ}$ (dot dashed) and 90$^{\circ}$ (lower solid). } 
\label{fig:incline2}
\end{figure}

Here we shall focus on a model from each of the two stellar evolution 
sequences which most closely approximate the average conditions of Achernar.
As one might expect from Figs\@. \ref{fig:surf} and \ref{fig:surf2}, the 
observed SED depends
on the inclination of the observer to the rotation axis.  
This is shown in Figs.~\ref{fig:incline1} and 
\ref{fig:incline2}, which shows the 
observed spectrum of models inclined at 0, 30, 60 and 90 degrees.  
The spectra shown in Fig.~\ref{fig:incline1} are based on an evolved
 6.5 M$_{\odot}$ 
model with uniform rotation on the zero-age main sequence (ZAMS) and a surface
equatorial velocity of v = 495km s$^{-1}$.  
This model has been evolved to a temperature and luminosity of T = 14510K, 
L = 3311 L$_{\odot}$, corresponding to the effective temperature and 
luminosity of $\alpha$ Eri \citep{code76}.  These observed parameters
were determined without considering the effects of rotation, and so are
only apparent parameters.
At this point, our model has an oblateness of only a/b = 1.19.  
Those in Fig.~\ref{fig:incline2} are based on an evolved 7.0 M$_{\odot}$ 
model, rotating on the ZAMS with a power law described by Eqn. \ref{eqn:power}.
The ZAMS surface equatorial velocity of this model is v = 430km s$^{-1}$.
The ratio of equatorial axis to polar axis is a/b = 1.32 on the ZAMS.
All spectra in this section are calculated assuming a distance of 40.0 pc to
Achernar, based on the OAO-2 data \citep{code76}.

\clearpage
\clearpage
\clearpage
\clearpage

We defined four passbands based on the variation of the spectra among the 
model atmosphere grid to generate color indices for evaluating the 
properties of these models.  The four passbands are: A: 1440-1460 
$\mbox{\AA}$, B: 1250-1280 $\mbox{\AA}$, C: 3100-3180 $\mbox{\AA}$ D: 1900-1940 
$\mbox{\AA}$.  The color indices we used are A-B, A-C and A-D.  
As a fourth color, we also calculated a Ly $\alpha$ index, taking the ratio 
of the flux at the bottom of the Ly $\alpha$ line (1210 $\mbox{\AA}$) to the 
flux at a point just redward of this line ($\sim$ 1240 $\mbox{\AA}$).  
We calculated
the color index for each one of the atmosphere models used to produce the
synthetic spectra,
which we then used to calibrate the color indices agains T$_{eff}$ and log 
{\it g}.  This allowed us to calculate apparent temperatures
for the synthetic SEDs.  This apparent temperature does not necessarily 
correspond to the physical temperature anywhere on the star, but gives an
effective average temperature, roughly corresponding to the observed 
temperature of the object.  
As a check on these 
inferred temperatures, we also calculated fits to the color-temperature data
for the other three color indices.  The results were quite similar for all
four indices, and suggest an uncertainty in these temperature estimates of 
$\pm$ 300 K.

\begin{figure}
\center
\plotone{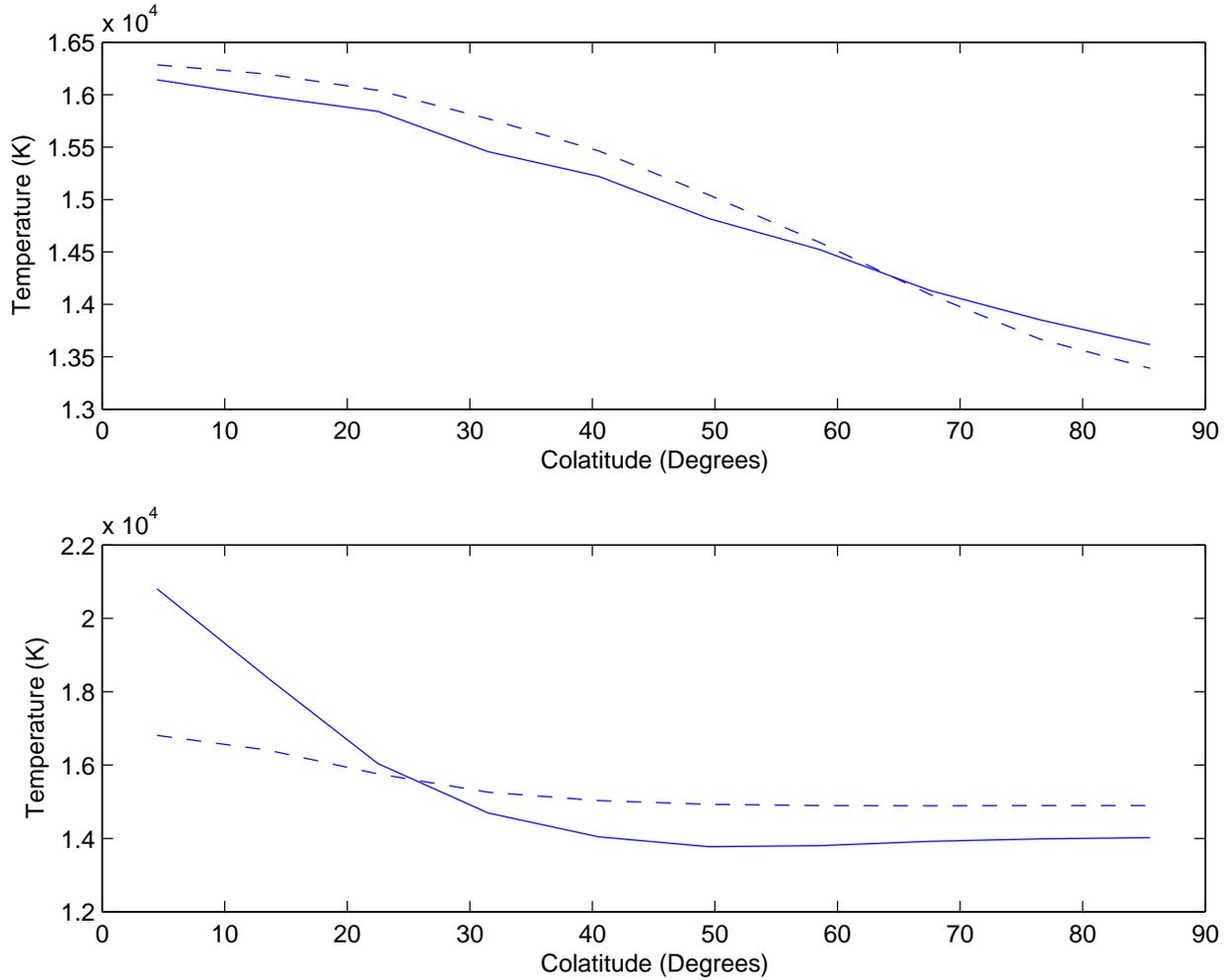}
\caption{Top: The ``observed'' effective temperature (solid) as a function of 
inclination for the 6.5 M$_{\odot}$ model.  
The dashed lines shows the calculated surface temperature as a function of 
colatitude for the same model.
Bottom:  Same as for top plot, but for the 7 M$_{\odot}$ model.  Both models
are evolved to approximately match the observed properties of Achernar.}
\label{fig:temp}
\end{figure}

Calculations were made every 10$^{\circ}$ of inclination.
Because the polar region of these oblate models is hotter than the equator, the
more pole on the star is,
the higher the apparent effective temperature of the star.  
In the spectra shown in Figs.~\ref{fig:incline1} and \ref{fig:incline2}, the 
inferred temperature difference between 0 and
90 degrees is between 2500 and 3000 K, depending on the details of the 
model.
This is illustrated in Fig.~\ref{fig:temp}, which shows the inferred 
effective temperature as a function of inclination for the two models described
above.  
\clearpage
\clearpage

Clearly, for more extreme angular momentum distributions, the temperature 
difference between the pole and the equator is larger.
The apparent effective temperatures for these
models ranges between 13000K and 18000K\@, and the luminosity ranges are 
correspondingly large.    
The inclinations that best correspond to the {\tt ROTORC} temperature of 14500K
are approximately 40$^{\circ}$ for the 7 M$_{\odot}$ model and 65$^{\circ}$ 
for the 6.5 M$_{\odot}$
model.  However, the inclinations required to match vsin{\it i} are 
90$^{\circ}$ and 82$^{\circ}$ respectively.
This suggests that the set of information contained in the observed L, 
T$_{eff}$ and v sin{\it i} data might be able to decouple the inclination,
but these limited calculations are insufficient to show either that this can 
be done or that the solution is unique.  Work in this area has been done by
\citet{maed70}, for example, and seems to indicate that the solution is 
indeed not unique.

\begin{figure}
\center
\plotone{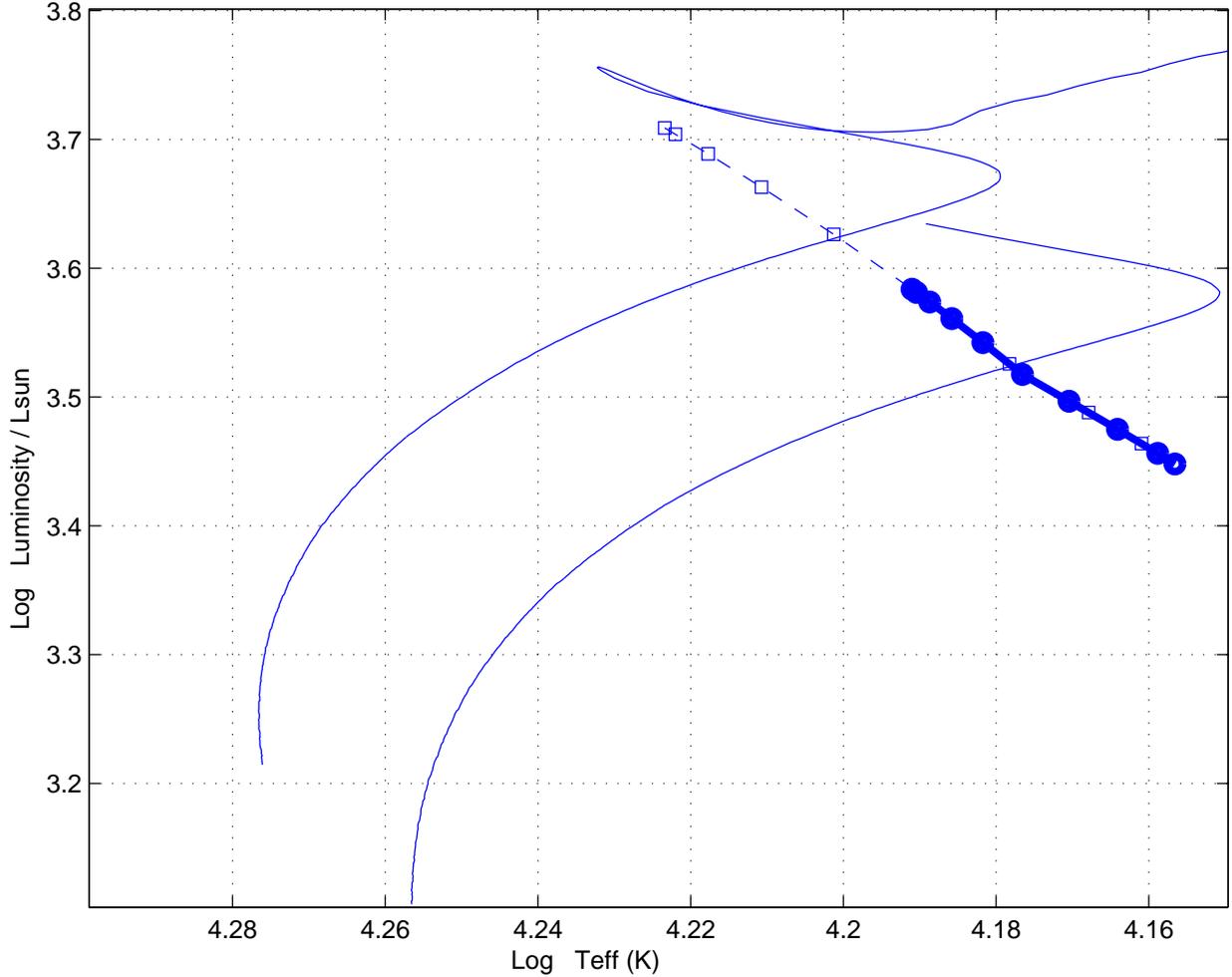}
\caption{The possible temperature and luminosity ranges of our models as functions of inclination.  The dashed line shows the values for the 7 M$_{\odot}$ model, while the solid bold line shows the values for the 6.5 M$_{\odot}$ model.  The points marked show the postion for every 10$^{\circ}$ of inclination (from left to right) 0$^{\circ}$ to 90$^{\circ}$ for the 6.5 M$_{\odot}$ ($\circ$) and 7 M$_{\odot}$ ($\square$) models respectively.  The evolutionary sequences for 7 M$_{\odot}$ and 6.5 M$_{\odot}$ uniformly rotating models are shown for reference.}
\label{fig:HR}
\end{figure}

The range of possible observationally determined temperature and luminosity 
for a given star is illustrated in Fig. \ref{fig:HR}.  
The range of possible values is centered
on the location in this case, as our {\tt ROTORC} temperature is an average.
A real observation of a single star results in a single point in the HR
diagram.  If the star is known to be rapidly rotating, this could result in
a huge uncertainty in its intrinsic position in the HR diagram.  
Without knowing the physical inclination of a rapidly rotating star, there is
no way to determine where on this curve the star actually lies.

\clearpage
\clearpage

\begin{figure}
\center
\plotone{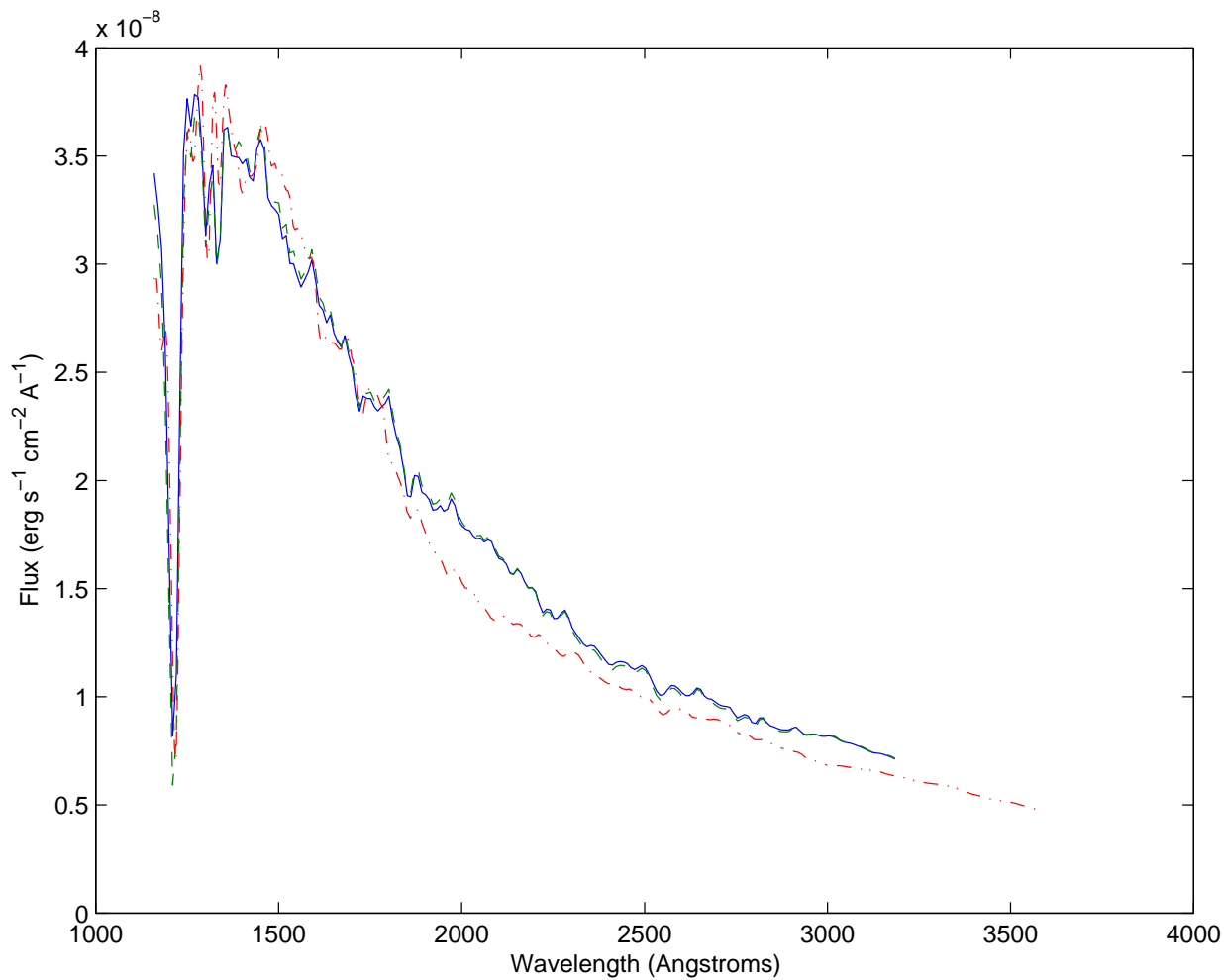}
\caption{The SED for a 6.5 M$_{\odot}$ model (solid) inclined at 80$^{\circ}$ 
and a 7 M$_{\odot}$ model (dashed) inclined at 84$^{\circ}$.  Although the 
structure of the two models is very different, there are only small 
differences in the SEDs.  
The 7 M$_{\odot}$ model, which has a larger surface temperature range, is
slightly closer to the observed spectrum of Achernar (dot-dashed).  This
suggests that models with very extreme surface variations, presumably 
requiring extreme differential rotation, could be used to reproduce the
observations.}
\label{fig:nodiff}
\end{figure}

It remains to be determined if the SED contains sufficient information to 
determine the pole to equator temperature range.  We compare the SEDs produced
by the 6.5 and 7 M$_{\odot}$ rotating models in Fig.~\ref{fig:nodiff}, along 
with the SED for Achernar, based on the OAO-2 data \citep{code79}.  
The properties for these three SEDs are summarized in Table 
\ref{table:achernar}.  The
inclinations of the two models were chosen to provide the best fit to each 
other at wavelengths greater than $\sim$ 1700 $\mbox{\AA}$.  The best fit
for the 6.5 M$_{\odot}$ model was chosen by visually matching the red tail
of the SED ($\lambda$ $>$ 2500 $\mbox{\AA}$) and was found to fit best at 
80$^{\circ}$.  We used 
linear interpolation to match the 7 M$_{\odot}$ model to the 6.5 
M$_{\odot}$ model, and found a match at 84$^{\circ}$.
To match the observed vsin{\it i} of Achernar, these models
must be inclined at 90$^{\circ}$ and 82$^{\circ}$ respectively.  

We find very few differences between the SEDs of the two models, and
neither provides a particularly good match to the observed SED of Achernar.
The two synthetic spectra give reasonably good matches throughout most of 
the tail region, beyond $\lambda$ $\sim$ 1700 $\mbox{\AA}$\@.  Near the peak of
the SED, the 7 M$_{\odot}$ model has a slightly higher flux than the 6.5 
M$_{\odot}$ model.  The observed spectrum of Achernar has even more flux in 
this region of the spectrum.  At this point in the evolution of the models, 
the 6.5 M$_{\odot}$ model is slightly more oblate, although the 7 M$_{\odot}$ 
model has a larger variation in surface temperature.  This suggests that to
successfully reproduce the observations would require even more extreme
differentially rotating models.
It may be possible to exploit differences that exist in the individual lines 
of these spectra \citep{col74,col79}, but this is beyond the scope of the 
present work.  

\clearpage
\clearpage

It is possible to reach an oblateness of 1.5 with an object rotating
uniformly very close to critical velocity, this does not explain the observed 
oblateness of Achernar.  Although an object uniformly rotating at critical 
velocity can reach
a high enough oblateness to match the observations, the resulting vsin{\it i} 
will not match the observed value for Achernar.  As the object must be viewed 
edge on to match the observed oblateness, i $\sim$
90$^{\circ}$.  This implies that the observed vsin{\it i} of 225 km s$^{-1}$
is the actual velocity of the object, yet this is clearly well below critical
rotation for this type of star.  As the star evolves along the main sequence,
the problem worsens.  There is no reason to believe the star maintains 
uniform rotation, and as the surface expands, the surface velocity will drop
and the star will become less oblate.  As Achernar appears to be an evolved
main sequence star, to have the observed oblateness at the observed 
vsin{\it i} requires differential rotation.
\clearpage
\begin{table}
\begin{center}
\caption{Properties of models compared to the observed 
properties of Achernar}
\begin{tabular}{||c|c|c|c|c|c||}
\hline
Model  &  T$_{eff}$ (K) & L (L$_{\odot}$)  &  v$_{eq}$ (km s$^{-1}$)  &  Inclination & a/b (observed)\\
\hline
\hline
Achernar  &  14510  &  3311   &  225$^1$  &  unknown &  1.56\\
\hline
6.5 M$_{\odot}$  &  14649  &  3377  &  223 &  82$^o$ &  1.20\\
\hline
7.0 M$_{\odot}$  &  14492  &  3752  &  208 &  90$^o$ &  1.17\\
\hline
\end{tabular}
\end{center}

\hspace{1cm}
$^1$  Observed v sin{\it i}.

\hspace{1cm}
{\tt Note:}  The observed oblateness of the models is given based on the angle of inclination
required to match the observed vsin{\it i} of Achernar.
\label{table:achernar}
\end{table}
\clearpage

\section{Conclusions}

We have calculated the internal structure and surface variation of models for
two rapidly rotating stellar evolution sequences using the 2D stellar 
evolution code {\tt ROTORC}.  One sequence was uniformly rotating on the ZAMS, the
other differentially rotating.  This evolution code allows us to directly 
model the surface variation in effective temperature
and gravity, which we can then compare with the predictions made by von 
Zeipel's law.  

We find our models are reasonably close to the predictions of von Zeipel's 
law, although there are some pronounced differences.  The difference 
predicted by our uniformly rotating model is largely a result of the 
equatorial flux.  {\tt ROTORC} predicts a much higher equatorial flux than
the von Zeipel model, which must be compensated by higher flux at the pole 
to keep the total
luminosity the same.  We believe this difference arises as a result of an
inherent contradiction in von Zeipel's law.  One of the fundamental 
assumptions of 
this model requires that the temperature be constant on equipotential surfaces.
The surface is also assumed to be an equipotential surface.  However, the 
effective temperature varies over the surface of the star.  We believe this
decoupling of the surface and effective temperatures gives rise to the
difference between the two models.
The differences are similar in form, although reduced in magnitude, when the
evolved model is compared.

For differentially rotating models, the situation is quite different and the
agreement with von Zeipel's law is not as good as for the uniformly rotating
models.  Our
models predict an appreciably higher temperature at the pole than the von 
Zeipel model.
We have performed several calculations to investigate the source of this
discrepancy, including varying the model
zoning and some details of the calculations in the convective core.  We found
that none of these changes had any significant effect on the temperature 
differences, leading us to suspect that the discrepancy is produced by some
aspect of our surface treatment.


Following the example of many previous studies, we have calculated the 
spectral energy distribution of a deformed star.  However, unlike previous
work, which relied on von Zeipel's law, our SEDs are based on the surface 
parameters obtained directly from 2D stellar structure models.  While our 
models are rotationally deformed, in principle
this method could be used on any type of deformed star, such as a companion in
a close binary.
This method is also valid over any spectral range and resolution, as long as
the appropriate model atmospheres and intensity grids can be produced.  
However, at higher resolution, Doppler effects would need to be included.
 
We find significant differences in the observed SED as a function of the 
inclination
of the rotation axis to the observer.  These differences could mean that the
effective temperature determined by an observer may have no relation to the
physically meaningful blackbody temperature of the star as a whole.  
By comparing the SEDs resulting from two different stellar structure models,
we have found that there are a few minor differences.  These are
not necessarily related to the oblateness of the model, but do seem to 
depend on the variation in surface temperature from pole to equator.  For these
models, the greater this variation, the more sharply peaked the resulting UV 
spectrum.

We have also attempted to find a match to the spectral energy distribution
of Achernar based on the OAO-2 observations \citet{code79}.  Of the synthetic
SEDs we have produced, the best matches are models inclined at 80$^{\circ}$
and 84$^{\circ}$, corresponding to the 6.5 and 7 M$_{\odot}$ models 
respectively.  These inclinations also correspond quite well to the 
inclinations required to match the observed vsin{\it i} of Achernar, 
82$^{\circ}$ and 90$^{\circ}$ respectively.  Unfortunately, our matches are 
far from perfect, particularly near the peak of the UV spectrum, near 1500 
$\mbox{\AA}$.  Neither of the underlying stellar models were as
oblate as the observations of \citet{dom03} indicate Achernar to be. 
If the increased oblateness also corresponds to an increase in the difference
in surface temperature from pole to equator, than it is possible that 
sufficiently differentially rotating models could reproduce the observations.
We expect to produce models with higher angular momentum distribution in the
near future.

\acknowledgements
This work was supported by a Discovery grant and a graduate scholarship from
Natural Sciences and Engineering Research Council of Canada (NSERC).  We would 
also like to thank
the Canada Foundation for Innovation and the Nova Scotia Research Innovation 
Trust for providing the computing facilities used in this project.

\end{document}